\documentclass[10pt, superscriptaddress,twocolumn]{revtex4}

\usepackage{upgreek,hyperref}
\usepackage{graphicx}

\begin{document}

\title{Development of integrated mode reformatting components for diffraction-limited spectroscopy}

\author{David G. MacLachlan}
\email{dgm4@hw.ac.uk}
\affiliation{SUPA, Institute of Photonics and Quantum Sciences, Heriot-Watt University, Edinburgh, EH14 4AS, UK}
\author{Robert J. Harris}
\affiliation{Department of Physics, University of Durham, South Road, Durham, DH1 3LE, UK}
\author{Debaditya Choudhury}
\affiliation{SUPA, Institute of Photonics and Quantum Sciences, Heriot-Watt University, Edinburgh, EH14 4AS, UK}
\author{Richard D. Simmonds}
\affiliation{Department of Engineering Science, University of Oxford, Parks Road, Oxford, OX1 3PJ, United Kingdom}
\author{Patrick S. Salter}
\affiliation{Department of Engineering Science, University of Oxford, Parks Road, Oxford, OX1 3PJ, United Kingdom}
\author{Martin J. Booth}
\affiliation{Department of Engineering Science, University of Oxford, Parks Road, Oxford, OX1 3PJ, United Kingdom}
\author{Jeremy R. Allington-Smith}
\affiliation{Department of Physics, University of Durham, South Road, Durham, DH1 3LE, UK}
\author{Robert R. Thomson}
\affiliation{SUPA, Institute of Photonics and Quantum Sciences, Heriot-Watt University, Edinburgh, EH14 4AS, UK}

\begin{abstract}
We present the results of our work on developing fully integrated devices (photonic dicers) for reformatting multimode light to a diffraction limited pseudo-slit. These devices can be used to couple a seeing-limited telescope point-spread-function to a spectrograph operating at the diffraction limit, thus enabling compact, high-resolution spectrographs that are free of modal-noise.
\end{abstract}

\maketitle

\section{Introduction}
The field of astrophotonics seeks to apply photonic technologies to astronomical instruments, with the aim of enabling improved performance \citep{Bland2009}. To harness the potential of photonics for astronomy, it is necessary to operate in the single-mode (SM) regime. In the case of ground based telescopes, the effect of atmospheric "seeing" imparts a rapidly changing aberration on the point-spread-function (PSF) preventing efficient coupling of a telescope Airy-pattern PSF to a single-mode-fiber (SMF). The PSF can be thought of as being composed of orthogonal spatial modes, the number of which can be approximated as:
\begin{equation}
M\approx(\pi\ \theta_{\rm{Focus}} D_{\rm{T}}/4\lambda)^2.
\label{Eq1}
\end{equation}
\noindent where $M$ is the number of modes that form the telescope PSF (counted as per polarization state), $\theta_{\rm{Focus}}$ is the angular size of the PSF in radians, $D_{\rm{T}}$ is the telescope diameter and $\lambda$ is the wavelength of the light \cite{Spaleniak:13, Harris2013}.

The phases and amplitudes of these modes change on millisecond timescales with atmospheric turbulence. In the context of coupling celestial light to optical fibers, the PSF can effectively be considered as being composed of a set of incoherent modes, only one of which on average can be coupled to an SMF. One approach to increase the coupling efficiency is to use adaptive optics (AO), which aims to reduce the contribution due to seeing in Eq. \ref{Eq1}. Future AO systems may enable coupling efficiencies in the near infrared approaching 100\% \cite{Jovanovic_SPIE}. The situation becomes considerably more difficult as either the wavelength of light is reduced and/or the size of the telescope aperture is increased, as revealed by Eq. \ref{Eq1}. Given the significant engineering challenges in developing AO systems for 8 m class telescopes, it is unlikely that AO systems will be able provide high SMF coupling efficiencies on extremely large telescopes (ELTs) with apertures $\geq$ 30 m, such as the European-ELT and Thirty Meter Telescope.

The photonic lantern is a guided-wave device that was developed to enable the efficient implementation of SM photonic function on seeing-limited telescopes. It is formed by a guided-wave transition between a multimode (MM) waveguide or fiber supporting M modes and a set of N individual single spatial modes. If the transition is sufficiently gradual, and N $\geq$ M, then the system can, in principle, couple the M incoherently excited modes at the MM input to the N output SMs with low loss.

To date, the highest performance photonic lanterns have been fabricated using optical-fiber-based techniques. Birks et al \cite{Birks12} recently demonstrated a MM-to-SM-to-MM transition based on photonic lanterns fabricated using a 120-core multicore fiber exhibiting a throughput loss of $<$ 0.5 dB. Photonic lanterns have also been fabricated using stack-and-draw techniques, and Noordegraaf et al \cite{ Noordegraaf:10} have demonstrated a 61 fiber based MM-to-SM-to-MM device with a  transmission loss $<$ 0.76 dB.

\begin{figure}
\centering
\includegraphics[width=0.9\linewidth]{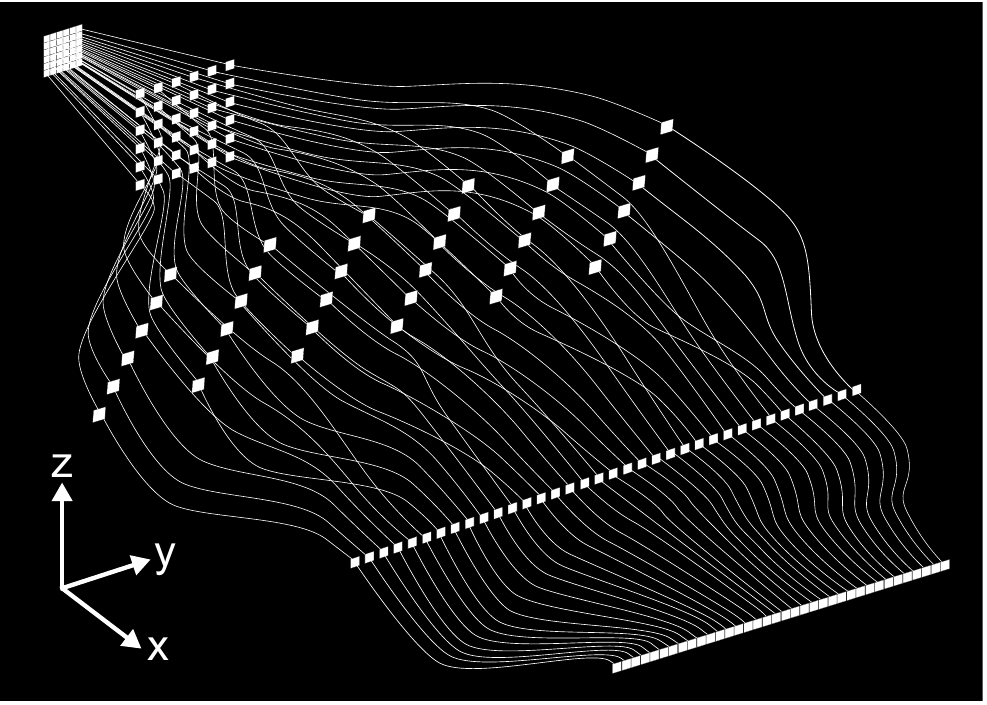}
\caption{3D sketch of the waveguide paths that were inscribed to create a PD with a MM input waveguide formed from a 6 x 6 close-packed array of SM waveguides. 
 }
\label{6x6schem}
\end{figure}

In 2011, we demonstrated \cite{Thomson11} that photonic lanterns could also be fabricated using a laser writing technique known as Ultrafast Laser Inscription (ULI), where focused ultrashort laser pulses are used to directly inscribe localized refractive index modifications inside a transparent dielectric material. We later demonstrated \cite{Harris} that ULI could be utilized to create a monolithic integrated device capable of reformatting a MM telescope PSF to a diffraction-limited pseudo-slit - a device that we named a "photonic dicer" (PD). Fig. \ref{6x6schem} shows a schematic diagram of a PD component, where it can be seen that the MM end of the device is formed from a 6 x 6 array of SM waveguides that are tightly stacked in the z- and y-axis (as seen in  Fig. \ref{6x6schem}). These waveguides split up along the x-axis via a photonic lantern transition into a two-dimensional array of loosely coupled SM waveguides. These waveguides are then routed through a series of transitions, to eventually form a planar waveguide that is SM across the waveguide (z-axis), and MM along the waveguide (y-axis). As originally proposed in the PIMMS (Photonic Integrated Multimode Micro-Spectrograph) instrument concept \cite{BlandHawthorn10}, the use of photonic lantern based devices for this purpose could revolutionize high-resolution astronomical spectroscopy, by enabling spectrographs that combine the high-efficiencies associated with a MM fiber-fed spectrograph, with the precision and stability of a SMF-fed spectrograph. In this paper, we present the results of detailed studies on the development of PD devices.

\section{Fabrication details}
The PD fabrication was performed using a 500 kHz train of 460 fs pulses at a wavelength of 1064 nm (Fianium Femtopower 1060fs laser). The pulse energy was set to 251 nJ and the polarization to circular.  The pulses were focused with a numerical aperture of 0.3 to a depth of $\approx$ 200 $\upmu$m below the surface of a borosilicate substrate (Corning EAGLE$^{2000}$). The substrate was mounted on high precision x-y-z crossed-roller bearing stages (Aerotech ANT) and translated through the laser focus at a speed of 8 mm.s$^{-1}$.

\begin{figure}[!htpb]
\centering
\includegraphics[width=0.9\linewidth]{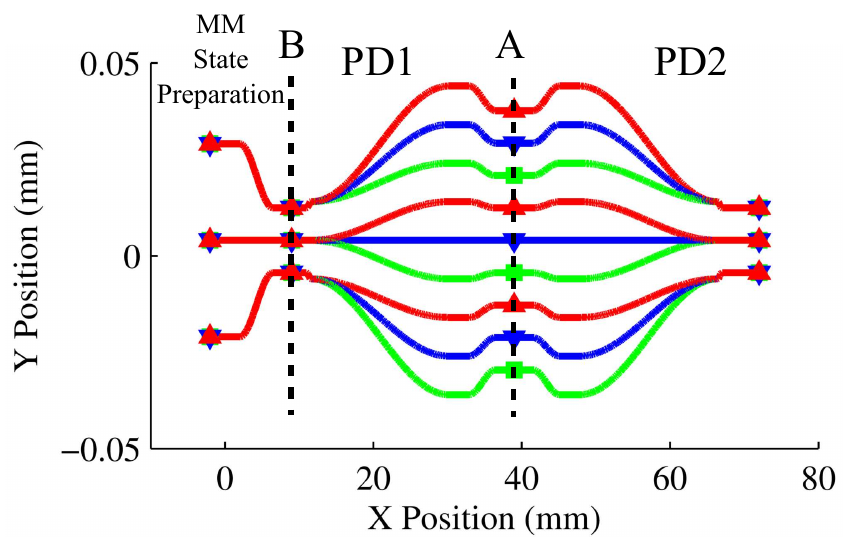}
	\caption{Schematic of the 3x3 PD showing the three sections, input, PD1 and PD2. Dashed lines indicate dice planes for cutback measurements. The colors represent different waveguide depths within the substrate at the MM state preparation.}
\label{3x3schem}
\end{figure}

\begin{figure}[!htpb]
\centering
\includegraphics[width=0.9\linewidth]{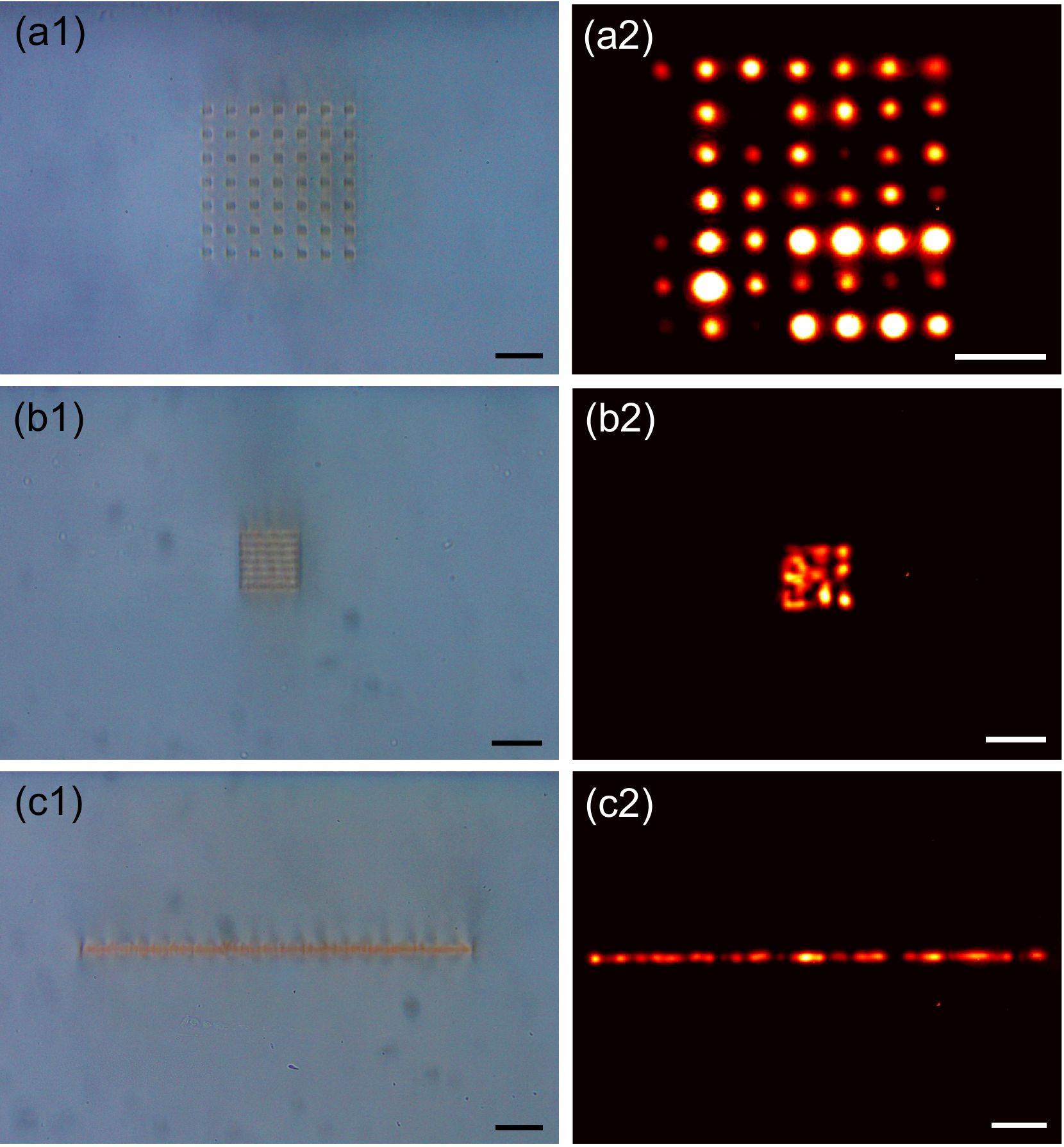}
	\caption{Images of the facets of (a1) the uncoupled SM input end (b1) the MM end and (c1) pseudo-slit end of one of the 7 x 7 PDs. Images in column (2) show near-field images of 1550 nm light emitted from the structures shown in (a), (b) and (c) respectively. The scale bar in each image represents 50 $\upmu$m.}
\label{facets}
\end{figure}

The cross section of the SM waveguides was controlled using the well-established multiscan technique \citep{Said2004}, which enables the fabrication of waveguides with an almost square cross section that can be positioned in close proximity to each other. Each SM waveguide was constructed by translating the substrate through the laser focus 21 times, with each scan laterally offset from the previous by 0.4 $\upmu$m. These square waveguides can then be stacked on top of, and adjacent to, each other with a center-to-center separation of 8.4 $\upmu$m to form either a large MM waveguide, or a pseudo-slit, that is SM across the slit (z-axis in Fig. \ref{6x6schem}) and MM along the slit (y-axis in Fig. \ref{6x6schem}). If the MM input to the device is formed by an K x K array of SM waveguides, we fabricated photonic dicers of six different "sizes", where K was varied between 2 and 7 in steps of 1. For each size of photonic dicer, we inscribed two 30 mm long devices (PD1 and PD2) back-to-back, seamlessly connected at their slit-ends. At one end of this structure, we also inscribed a 10 mm long "MM-state-preparation" section, a schematic of the design for a 3x3 full device is shown in Fig. \ref{3x3schem}. The state-preparation input consisted of an K x K array of SM waveguides separated by a center-to-center distance of 25 $\upmu$m that approach each other along the length of the section to form a MM waveguide matched to the MM end of PD1. Microscope images of the facets of the input, MM and pseudo-slit ends of a 7 x 7 PD, together with near-field images of 1550 nm light emitted from these structures are shown in Fig. \ref{facets}.

\section{Device characterization}
\subsection{Third harmonic imaging}
\begin{figure}[!h]
\centering
\includegraphics[width=0.9\linewidth]{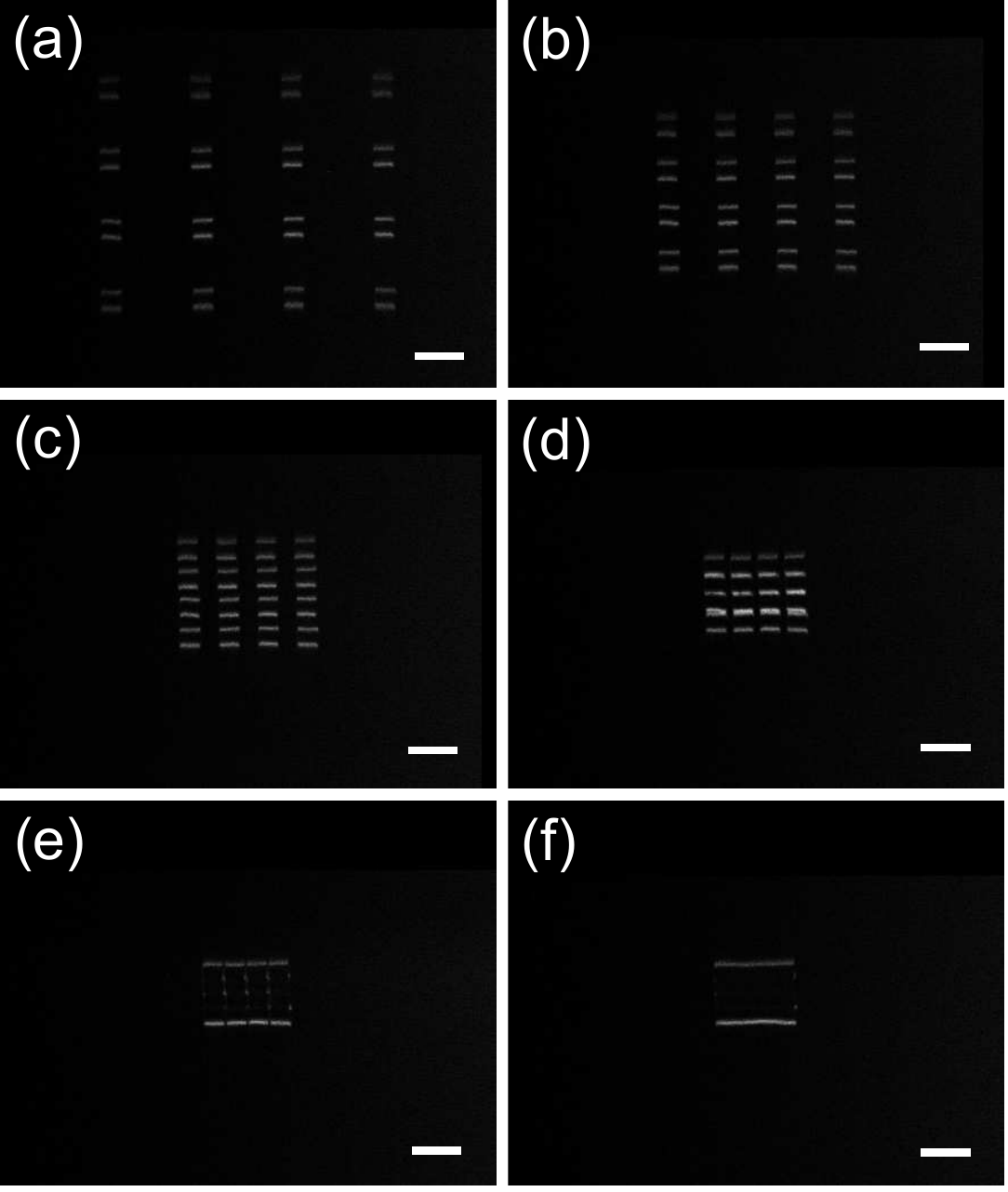}
\caption{Cross-sectional images of a ULI fabricated lantern, obtained using a third-harmonic-generation microscope. Bright areas indicate a material boundary. Fig. (a) is furthest from the multimode end of the device, and Fig. (f) is within the multimode end of the device. Scale bars indicate 20 $\upmu$m.}
\label{thg}
\end{figure}

	We used nonlinear third-harmonic-generation (THG) to obtain cross-sectional images through a ULI fabricated lantern manufactured on a different system, though with similar parameters to those used in the manufacture of the photonic dicers with full details reported in \cite{Thomson11}. The system used for the THG imaging was based on a Cr:forsterite laser producing a 76 MHz train of 65 fs pulses at a wavelength of 1235 nm, and is similar to the system reported in \cite{Jesacher09} which has previously been used to image ULI fabricated optical waveguides with 3D resolution \citep{Marshall11}. The THG signal is only generated when the pulses are focused at a boundary between different materials, for example between the modified and unmodified regions within the substrate. The images shown in Fig. \ref{thg} are therefore highly informative. The "double horizontal line" structure of the SM waveguides in images (a) to (d) strongly suggest that the SM waveguides exhibit a close to top-hat refractive index profile, as has been confirmed previously in multiscan fabricated waveguides \citep{Nasu05}. A much weaker THG signal is obtained at interfaces along the optical axis than those normal to the optical axis, hence the double horizontal line appearance for a square area of modified material. Images (e) and (f) demonstrate that as the SM waveguides approach each other, the THG signal becomes progressively weaker until almost no contrast in the image can be seen within the MM waveguide in Fig. \ref{thg} f. This is exactly the evolution one would require for an efficient and adiabatic MM-to-SM transition.

\subsection{Throughput characterization}
Coherent MM states were generated to test the throughput of the PDs by butt-coupling SMF28 fiber into each of the SM input waveguides on the state preparation section, and the output from the MM end of PD2 measured by free space coupling to a photodiode (Measurement Set 1). PD2 was then diced from the device (along line A in Fig. \ref{3x3schem}), and the output of the PD1 slit was measured (Measurement Set 2). Finally PD1 was diced from the device (along line B in Fig. \ref{3x3schem}), and the measurement repeated for light emitted from the MM end of the state preparation section (Measurement Set 3). These measurements effectively form a series of cutback measurements, where differences in signal between Measurements 1 and 2 provide the throughput losses of PD2 and the differences in signal between Measurements 2 and 3 provide the throughput losses of PD1.

\begin{figure}[!h]
\centering
\includegraphics[width=0.9\linewidth]{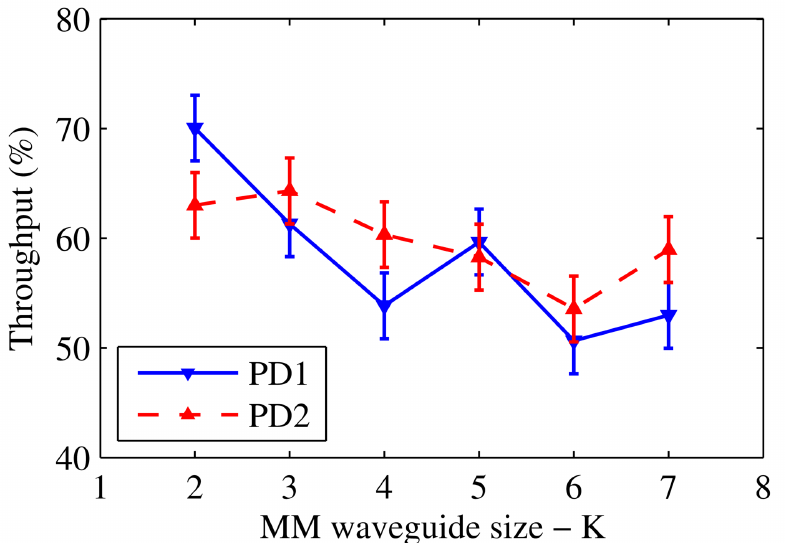}
\caption{Average throughputs measured for PD1 (MM-to-slit) and PD2 (slit-to-MM) for each K by K photonic dicer.}
\label{avthru}
\end{figure}

\begin{figure}
\centering
\includegraphics[width=\linewidth]{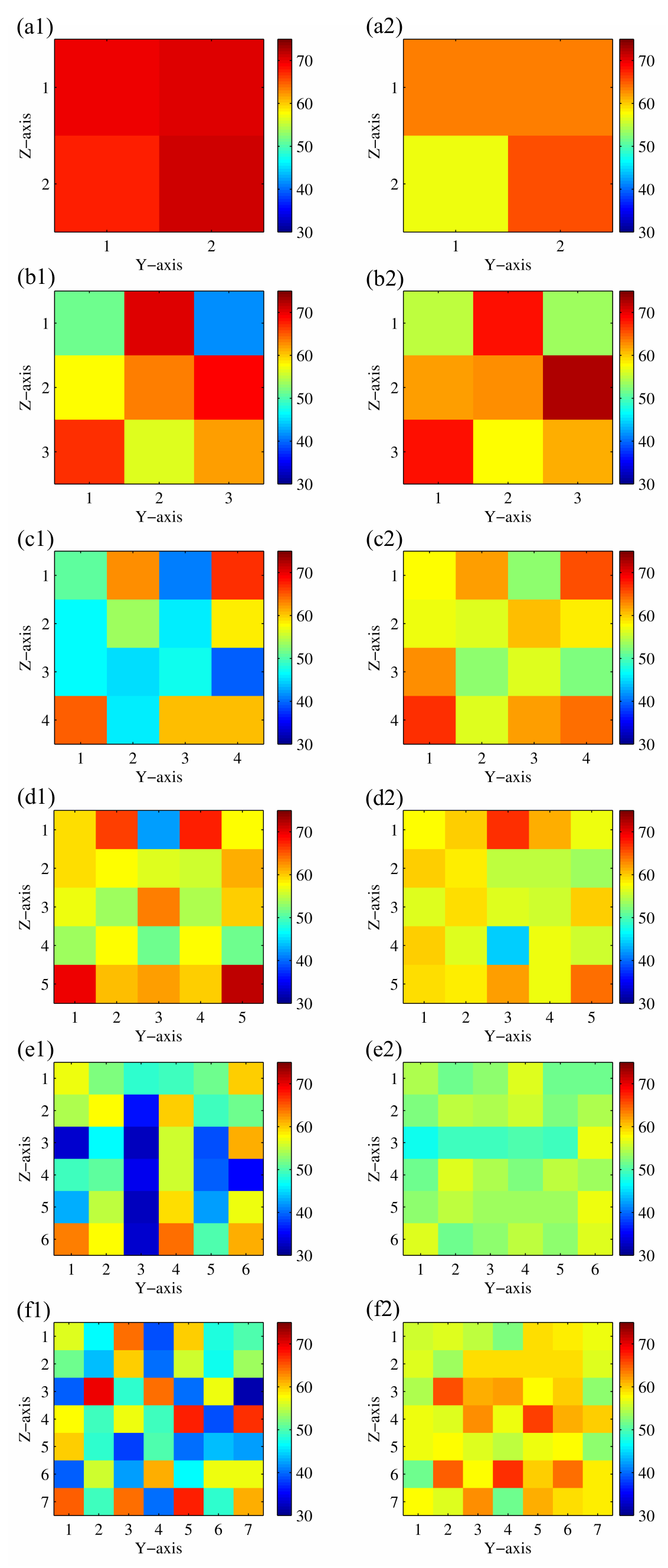}
\caption{Throughputs measured for PD1 (Left column) and PD2 (Right column). The size of the PD under test increases from (a) to (f). For each set of measurements, each small square represents the percentage throughput measured when injecting 1550 nm light into an input waveguide on the MM state preparation device.
 }
\label{thru}
\end{figure}

The throughput of each photonic dicer was measured at 1550 nm for horizontal and vertical polarizations and the average determined. The 2 x 2 PD structure was also investigated for 1500 and 1580 nm light, in both cases the throughput was found to be similar to that measured for 1550 nm light. The individual waveguides used to construct the devices were measured to be single mode throughout the 1320 to 1580 nm region, so we expect that the PD devices will work with a similar performance across this range. The results of the throughput measurements are presented in Figs. \ref{avthru} and \ref{thru}. The average throughput is seen to decrease steadily as the size of the PD increases (Fig. \ref{avthru}) as the waveguides within larger PDs must perform tighter bends, therefore exhibiting higher radiation losses than those within smaller devices. It is also interesting to note that the average throughputs for PD1 and PD2 are very close to each other, despite the fact that the light was propagating in the MM-to-slit direction for PD1, and the slit-to-MM direction for PD2. This observation would tend to indicate that the photonic-lantern and slit-forming transitions are close to adiabatic, with the number of modes being supported along these transitions remaining roughly constant. A further interesting feature to note is that even though the average throughput measured for PD1 and PD2 is approximately the same, the spread of throughputs is significantly decreased for PD2 (Fig. \ref{thru}). This observation is indicative of mode scrambling due to optical path length differences within the PD, as discussed previously in \citep{Birks12} and \cite{Birks15}.  

\section{Conclusions}
We have presented a body of work on the development of PD devices using ULI. We have used THG microscopy to image a ULI fabricated photonic lantern transition. The results of this clearly demonstrate that the multiscan technique enables the fabrication of a well-controlled transition, where individual step index waveguides can be slowly combined to create a single, near-step-index MM waveguide. We have also performed detailed throughput characterization experiments for PD devices of different sizes. We observed that the throughput steadily reduced as the size of the PD increased, a property that we attribute to increased radiation losses as the routing waveguides are forced to perform increasingly severe bends as the PD size increases. Further throughput and performance tests of a 6x6 PD device were performed on-sky and are presented in \cite{Harris}.

\label{sec:examples}

\section*{Funding Information}

Science and Technology Facilities Council (STFC) (ST/K00235X/1) (ST/I505656/1) (ST/K000861/1). OPTICON Research Infrastructure for Optical/IR astronomy (EU-FP7 226604). Engineering and Physical Sciences Research Council (EPSRC) (EP/E055818/1).

\section*{Acknowledgments}
We thank Dr Graeme Brown (Optoscribe Ltd) for assistance with the stage programming.

\end{document}